# Coordinated Dynamic Operation of Integrated Electrolyzer–Compressor Systems

Amin Salehi, *Student Member, IEEE*, Janne Seppänen, *Senior Member, IEEE*, Mahdi Pourakbari-Kasmaei, *Senior Member, IEEE*

***Abstract*— The increasing interaction between power and hydrogen sectors highlights the importance of coordinated operation of electrolyzers and electric-driven compressor stations (EDCSs). This becomes particularly of higher importance under transient disturbances. However, coordinated dynamic interactions of these coupled subsystems remain largely unexplored. This article addresses such gap by developing a dynamic model for an integrated electrolyzer-EDCS system and designing appropriate PID control schemes to address the potential disturbances affecting either component. To this end, linearized models of the electrolyzer and EDCS are first derived to enable systematic controller design. Then, two PID controllers, representing conservative and fast-tracking designs, are developed to coordinate the system response. The developed coordinated model is examined and verified under four different cases. The results demonstrate the effectiveness of the proposed model under disturbances from the compressor driver or the electrolyzer. Controlling the electrolyzer flow in response to EDCS disturbances coordinates system dynamics and mitigates undesirable transient fluctuations. Conversely, under electrolyzer disturbances, regulating the EDCS torque eliminates inconsistent responses in pressure, flow, and rotational speed, while preventing hazardous transient undershoots and overshoots. Overall, the proposed framework guarantees transient stability and operational reliability of the integrated electrolyzer-EDCS system.**

***Index Terms*— Power and hydrogen integration, electrolyzer, compressor station, dynamic modeling, transient stability.**

## I. INTRODUCTION

THE widespread adoption of renewable energy resources and power-to-hydrogen technologies has facilitated greater integration of power and hydrogen ($H_2$) networks [1], [2], [3]. In such integrated systems, electrolyzers and electric-driven compressor stations (EDCSs) serve as the key coupling components between the two sectors [4], [5]. Electrolyzers can convert surplus renewable electricity into $H_2$, providing grid flexibility and supporting the decarbonization of heavy industry and transportation sectors [6], [7]. Subsequently, compressors are deployed to raise $H_2$ pressure to required level of transmission pipelines, refueling stations, and storage facilities [4], [8]. As the $H_2$ sector grows, the interdependence between the power and $H_2$ sectors becomes stronger, necessitating coordinated dynamic operation of joint electrolyzers and EDCSs, a topic that remains unexplored in the existing literature. Such coordination becomes particularly crucial as the system's stability needs to be guaranteed under disturbances affecting either the electrolyzer or EDCS in the integrated system.

The dynamic behavior of electrolyzers under variable electrical conditions has been broadly studied in the literature. The authors in [9], [10], [11], [12], [13] developed dynamic modeling frameworks for utility-scale electrolyzers and showed their capability to provide fast frequency response and ancillary services in renewable-rich power systems. Control-oriented modeling and control of electrolyzers have also been widely investigated, highlighting the importance of dynamic models for designing effective operational and supervisory control strategies [14], [15], [16], [17], [18]. Furthermore, experimental and simulation studies have analyzed the response of electrolyzers to renewable fluctuations and their integration into power systems [19], [20]. Collectively, these studies confirm the ability of electrolyzers to respond dynamically to grid-side events. Nevertheless, their interaction with downstream $H_2$ compression stations remains insufficiently explored.

EDCSs are another key coupling point between the power and $H_2$ networks. Their motors are supplied by the power grid, whereas their operating conditions depend on $H_2$ flow and pressure dynamics. These components are connected to the electrolyzer plants to increase the pressure level of the produced $H_2$. Hence, two types of disturbances could impact their dynamic operation [4]. Variations in EDCS driver torque can affect compressor rotational speed, mass flow rate, suction pressure, and discharge pressure [4], [21]. Moreover, electrolyzer flow rate fluctuations caused by power grid disturbances can alter the EDCS suction flow, thereby affecting its operation [4]. Several studies have therefore focused on designing appropriate control schemes for the EDCS to stabilize its operation [22], [23], [24], [25], [26]. However, the coordinated dynamic operation of the electrolyzer and EDCS has not been thoroughly analyzed. In large-scale integrated power–hydrogen systems, such dynamic interaction prevents lags or mismatches in the production–consumption chain, where electrical and hydrogen components are strongly interdependent.

Despite the aforementioned developments, existing studies mainly consider electrolyzer flexibility, compressor control, or power-hydrogen coordination as separate problems. In particular, the bidirectional dynamic coordination between the electrolyzer and EDCS operation during disturbances has not been sufficiently investigated. This issue becomes critical when a disturbance reduces the EDCS driver torque without a corresponding adjustment in electrolyzer $H_2$ production, thereby creating a supply-compression mismatch that may lead to undesirable pressure and flow transients [27]. Similarly, if the electrolyzer output changes while the EDCS driver torque remains unchanged, the compressor may exhibit inconsistent speed, flow, and pressure responses. Therefore, a coordinated control mechanism is required to regulate the electrolyzer flow rate in response to disturbances in the EDCS and, conversely, to regulate the EDCS torque in response to electrolyzer disturbances.

This paper develops a coordinated control framework for an integrated electrolyzer-EDCS system. The electrolyzer is represented by an already existing linear dynamic model relating stack voltage to $H_2$ production flow, whereas the EDCS is modeled using nonlinear pressure, flow, and mechanical speed dynamics. The EDCS model is linearized to enable controller design. Based on the linearized models, PID controllers are designed using pole placement and polynomial matching. The proposed framework incorporates two complementary control actions: 1) regulation of electrolyzer flow rate following a disturbance in the EDCS driver, and 2) regulation of the EDCS driver torque after a disturbance in electrolyzer $H_2$ production. Accordingly, the main contributions of this work are summarized as follows.

1) Developing an integrated dynamic model of the coupled electrolyzer-EDCS system to capture the interactions between $H_2$ production and compression processes. Thus, the model fills the gap by relating the $H_2$ flow rate produced by the electrolyzer to the compressor flow, suction and discharge pressures, rotational speed, and driver torque.
2) Developing a linearized dynamic model for the EDCS and introducing a linear dynamic model for the electrolyzer. The relative state space representations are then formulated to facilitate systematic controller design.
3) Designing conservative and fast-tracking PID controllers for bidirectional coordination: a) the electrolyzer control unit regulates $H_2$ flow rate when the EDCS driver is imposed by a driver torque disturbance. This control action coordinates $H_2$ production with the compression system and mitigates undesirable pressure and flow transients; and b) the EDCS control unit adjusts driver torque in response to disturbances in the output flow of the electrolyzer. The proposed control strategy eliminates inconsistent responses in rotational speed, flow, and pressure of the EDCS. Consequently, it prevents the compressor from approaching the choke region under the disturbances.

## II. Control Concept and Objective in a Coupled Electrolyzer-Compressor System

An integrated renewable-electrolyzer-EDCS system is depicted in Fig. 1 as an example for the $H_2$ transmission case. Under normal operating conditions, the electrolyzer produces $H_2$ at a specified pressure. Then, the EDCS is employed to raise the pressure of the produced $H_2$ for transmission along the pipeline. This is necessary to fulfill network requirements, such as maintaining specific pressure levels and meeting certain gas demands [8]. Aside from normal operating conditions, two commonly occurring contingencies in such a system are loss of motor torque in the EDCS and $H_2$ flow-rate fluctuations in the electrolyzer [4], [28].

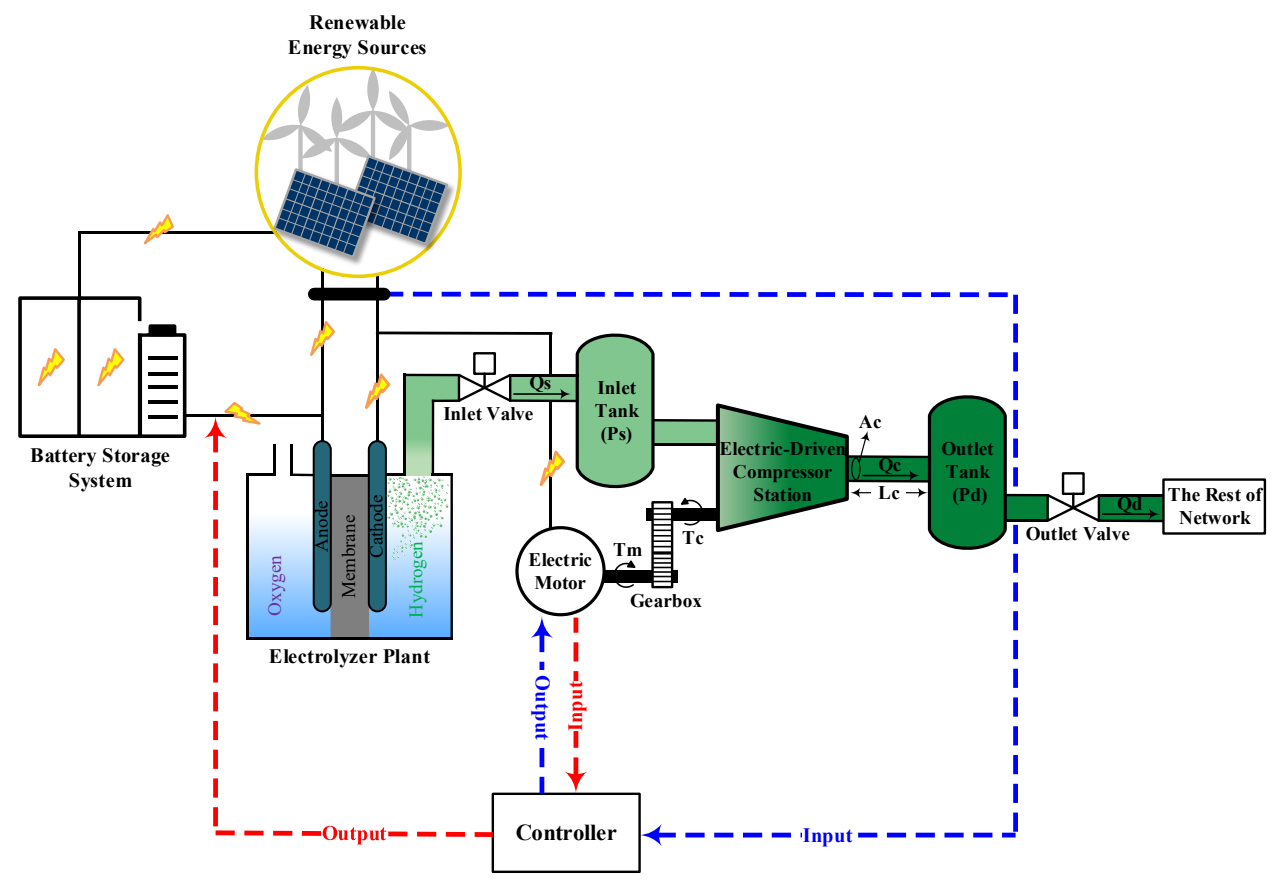

**Fig. 1.** Schematic diagram of an integrated electrolyzer-EDCS system.

### A. Contingency on the Compressor

The EDCS might experience a disturbance from the power grid side, resulting in torque fluctuations in its electric motor. In this situation, the electrolyzer's output flow rate should be controlled to mitigate EDCS fluctuations. This means that the EDCS suction flow rate would be controlled in response to its motor torque fluctuations. This will prevent the compressor from damaging and possible downstream load variations. Control signals corresponding to this situation are drawn with dashed red lines in Fig. 1. After sensing torque variations, a control command needs to be sent to the electrolyzer to adjust its output flow. The closed-loop block diagram of the control system in this case is shown in Fig. 2, with the plant being the electrolyzer. $Q_E$ and $Q_{E\text{-ref}}$ are the electrolyzer flow rate and its reference flow rate, respectively.

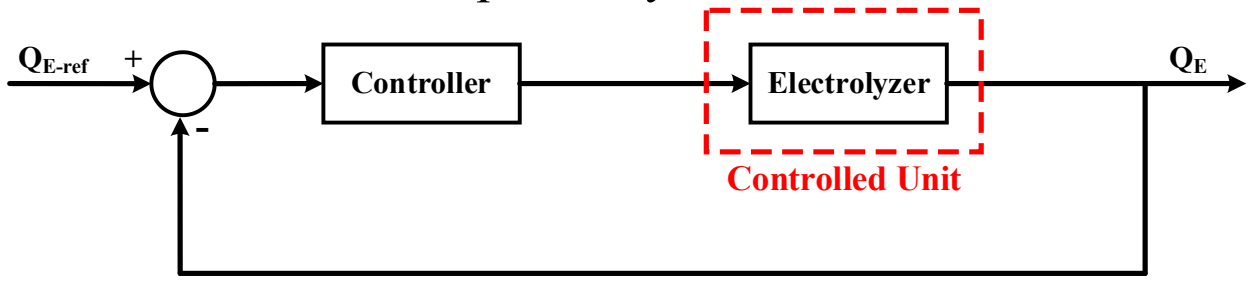

**Fig. 2.** Closed-loop block diagram of the control system when the EDCS is experiencing power grid disturbances.

### B. Contingency on the Electrolyzer

A power grid disturbance could also trigger flow fluctuations in the produced $H_2$ flow by the electrolyzer plant. In such circumstances, the motor torque of the EDCS should be regulated to manage $H_2$ fluctuations entering the compressor. That is the motor torque of the EDCS should be controlled in response to its suction flow fluctuations. This prevents equipment damaging and possible downstream load variations. Control signals for this case are shown with dashed blue lines in Fig. 1. After sensing voltage variations in the electrolyzer, a control command needs to be sent to the EDCS to regulate its driver torque. The closed-loop block diagram of the control system in this case is presented in Fig. 3, where the plant is the EDCS. $Q_C$ and $Q_{C\text{-ref}}$ are the compressor's flow rate and its reference flow rate, respectively.

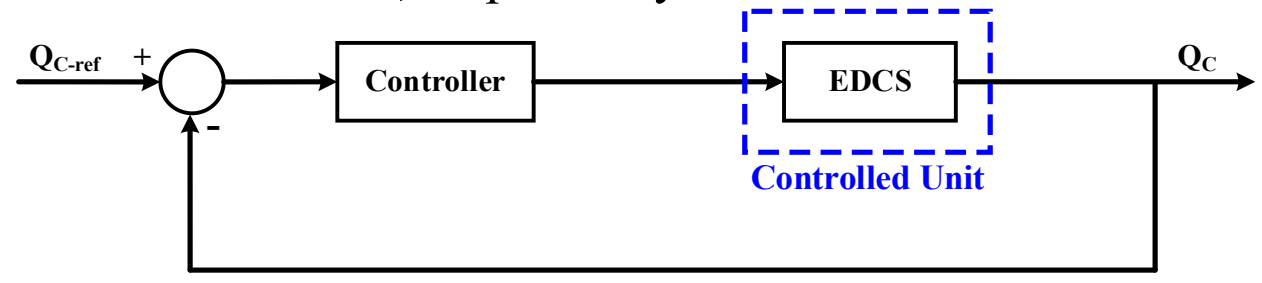

**Fig. 3.** Closed-loop block diagram of the control system when the electrolyzer is experiencing power grid disturbances.

## III. Controller Design for the Proposed System

To enable systematic controller synthesis, we start with the dynamic models of the electrolyzer and EDCS. For nonlinear models, all the nonlinearities will be first linearized around a nominal operating point corresponding to steady-state operation. Then, the linear state-space representations are obtained using Jacobian linearization. Finally, two PID controllers are designed based on the resulting linear models.

### *A. Linear State-Space Representation of the Electrolyzer System and Its PID Controller Design*

A linear dynamic model can be presented for both alkaline and proton exchange membrane electrolyzers based on the study performed in [29]. This model is represented by a second-order transfer function (1), in which the electrolyzer stack current is calculated from the electrolyzer stack DC voltage. Then, the electrolyzer current drives the electrolysis process, leading to $H_2$ production. As presented in (2), the $H_2$ flow rate is proportional to the current [30].

$$G_{\mathrm{E}}(s)=\frac{I_{\mathrm{E}}(s)}{V_{\mathrm{E}}(s)}=\frac{\alpha\beta}{(s+\alpha)(s+\beta)} \tag{1}$$

$$\frac{Q_{\mathrm{E}}(s)}{I_{\mathrm{E}}(s)}=\frac{\xi\eta_f N_s N_p}{2F} \tag{2}$$

where, $V_{\mathrm{E}}(s)$ and $I_{\mathrm{E}}(s)$ are the Laplace transforms of the voltage and current of the electrolyzer stack, respectively; α and β ($\alpha$ , $\beta > 0$) represent the poles of the stack transfer function; $Q_{\mathrm{E}}$ is the $H_2$ production rate in kg/s, ξ is the coefficient converting mol/s to kg/s, $\eta_f$ demonstrates the faraday efficiency; F is the Faraday's constant; and, $N_s$ and $N_p$ denote the number of stack cells in series and parallel, respectively. Combining (1) and (2), the relation between $Q_{\mathrm{E}}$ and $V_{\mathrm{E}}$ is presented in (3).

$$G_{\mathrm{E}}(s)=\frac{Q_{\mathrm{E}}(s)}{V_{\mathrm{E}}(s)}=\frac{\psi}{s^2+(\alpha+\beta)s+\alpha\beta} \tag{3}$$

where, ψ is the coefficient equal to $\frac{\xi\eta_f N_s N_p\alpha\beta}{2F}$.

Let the input be $u = V_{\mathrm{E}}$ and the output be $y = Q_{\mathrm{E}}$. Then, a standard controllable state-space realization for the electrolyzer system can be demonstrated as follows.

$$\dot{x}=\begin{bmatrix}0 & 1\\ -\alpha\beta & -(\alpha+\beta)\end{bmatrix}x+\begin{bmatrix}0\\1\end{bmatrix}V_{\mathrm{E}}$$

$$y=Q_E \tag{4}$$

Based on the closed-loop block diagram in Fig. 2, the tracking error is $e(t) = Q_{\mathrm{E\text{-}ref}}(t) - Q_{\mathrm{E}}(t)$. Also, standard PID representation and the closed-loop characteristic are shown in (5) and (6), respectively.

$$C_{\mathrm{PID}}(s)=K_p+\frac{K_i}{s}+K_d s=\frac{K_d s^2+K_p s+K_i}{s} \tag{5}$$

$$\Delta_{\mathrm{cl}}(s)=1+C_{\mathrm{PID}}(s)G_{\mathrm{E}}(s)=0 \tag{6}$$

Then, the closed-loop characteristic polynomial can be obtained by substituting (3) and (5) in (6). This is shown in (7).

$$\Delta_{\mathrm{cl}}(s)=s^3+(\alpha+\beta+\psi K_d)s^2+(\alpha\beta+\psi K_p)s+\psi K_i=0 \tag{7}$$

Equation (7) is the key to PID design by pole placement and polynomial matching. To this end, we pick a desired stable third-order polynomial, then match coefficients. Without loss of generality, the desired closed-loop poles can be chosen such as $s_{1,2,3} = p_1, p_2, p_3$ ($p_1, p_2, p_3 < 0$). Then, matching the desired polynomial with the closed-loop characteristic polynomial ($\Delta_{\mathrm{d}}(s) = \Delta_{\mathrm{cl}}(s)$), will determine PID gains as shown in (8).

$$\begin{aligned}&s^3+(\alpha+\beta+\psi K_d)s^2+(\alpha\beta+\psi K_p)s+\psi K_i=\\&s^3+(p_1+p_2+p_3)s^2+(p_1p_2+p_1p_3+p_2p_3)s+p_1p_2p_3\end{aligned} \tag{8}$$

where, PID gains are derived as follows.

$$\begin{aligned}K_p&=\frac{(p_1p_2+p_1p_3+p_2p_3)-\alpha\beta}{\psi}\\K_i&=\frac{p_1p_2p_3}{\psi}\\K_d&=\frac{(p_1+p_2+p_3)-(\alpha+\beta)}{\psi}\end{aligned} \tag{9}$$

Finally, based on the closed-loop performance objectives (e.g., fast reference tracking and conservative response), the desired poles are selected and PID gains are calculated accordingly. For the conservative design, the dominant poles are placed closer to the imaginary axis, resulting in a lower closed-loop bandwidth and slower response. In contrast, the fast-tracking design uses poles farther into the left-half plane, resulting in a higher bandwidth and faster nominal tracking. Both designs are investigated via simulations in Section V.

### *B. Linear State-Space Representation of the EDCS and Its PID Controller Design*

The EDCS dynamics are described by the following nonlinear differential equations [4], [27].

$$\dot{P}_s=\frac{a^2}{V_s}(Q_s-Q_c) \tag{10}$$

$$\dot{P}_d=\frac{a^2}{V_d}(Q_c-Q_d) \tag{11}$$

$$\dot{Q}_c=\frac{A_c}{L_c}(\Pi_c(\omega_c,Q_c)P_s-P_d) \tag{12}$$

$$\dot{\omega}_c=\frac{1}{J}(T_m-T_c-k_l\omega_c^2) \tag{13}$$

where, $P_{\mathrm{s}}$ and $P_{\mathrm{d}}$ are the inlet and outlet pressures at the compressor's suction and discharge sides, respectively. $Q_{\mathrm{s}}$, $Q_{\mathrm{c}}$, and $Q_{\mathrm{d}}$ are the inlet flow, the compressor flow, and the outlet flow, respectively; $a$ is the speed of sound, and $V_s$, $V_d$, $A_c$, and $L_c$ are the inlet tank volumes, outlet tank volumes, the cross section of the piping, and the duct length after the compressor, respectively; $\Pi_c$ *($\omega_c$, $Q_c$)* is a fitted polynomial map, presenting the pressure ratio ($P_d/P_s$); $\omega_c$ is the rotational speed of the compressor, J is the inertia of the mechanical system, $T_m$ is the driver torque, $T_c$ is the compressor torque, and $k_l$ is the rotating losses coefficient. Inlet and outlet flows, compressor torque, and the polynomial map can be modeled with (14)-(17), respectively.

$$Q_s=k_s u_{in}^3\sqrt{P_{in}-P_s} \tag{14}$$

$$Q_d=k_d u_{out}^3\sqrt{P_d-P_{out}} \tag{15}$$

$$T_c=k_c\omega_c Q_c \tag{16}$$

$$\Pi_c(\omega_c,Q_c)=\sum_{i=0}^{3}\sum_{j=0}^{3}\alpha_{ij}\omega_c^jQ_c^i \tag{17}$$

where, $k_s$ and $k_d$ are the inlet and outlet valve gains, $u_{in}$ and $u_{out}$ are the extent of the inlet and outlet valve openings, and $P_{in}$ and $P_{out}$ are the boundary condition associated with the inlet and outlet headers, respectively; $k_c$ is the friction constant and $\alpha_{ij}$ is a 16-dimensional parameter vector determined by the polynomial approximation. It is worth mentioning that $T_m$ is the input to the EDCS, and the $H_2$ generated by the electrolyzer serves as the inlet flow to the EDCS (i.e., $Q_E = Q_s$).

Unlike the electrolyzer, the EDCS dynamics include many nonlinear terms. Thus, linearization around steady-state operating points is required to design a proper PID controller. Let's consider the state and input vectors as $x = [P_s, P_d, Q_c, \omega_c]^T$ and $u = T_m$, respectively. The equilibrium points are also defined as $x_0 = [P_{s0}, P_{d0}, Q_{c0}, \omega_{c0}]^T$ and $u_0 = T_{m0}$. Compressor flow ($Q_c$) is considered the system output. The system can be generally written as $\dot{x} = f(x, u)$ and $y = h(x, u)$. Then, linearizing around an operating point ($x_0$, $u_0$) gives (18) and (19).

$$\delta\dot{x} = A\delta x + B\delta u \tag{18}$$

$$\delta y = C\delta x + D\delta u \tag{19}$$

where,

$$A=\frac{\partial f}{\partial x}\bigg|_{(x_0,u_0)}, \qquad B=\frac{\partial f}{\partial u}\bigg|_{(x_0,u_0)}$$
$$C=\frac{\partial h}{\partial x}\bigg|_{(x_0,u_0)}, \qquad D=\frac{\partial h}{\partial u}\bigg|_{(x_0,u_0)} \tag{20}$$

Let's start with matrix $A$. If $\dot{P}_s$ is considered as $f_1$, then:

$$\frac{\partial f_1}{\partial P_s}\bigg|_{(x_0,u_0)}=\frac{-a^2k_su_{in}^3}{2V_s\sqrt{P_{in}-P_{s_0}}}=a_{11}$$
$$\frac{\partial f_1}{\partial Q_c}\bigg|_{(x_0,u_0)}=-\frac{a^2}{V_s}=a_{13} \tag{21}$$

All other partials in the first row of $A$ are zero. If $\dot{P}_d$ is considered as $f_2$, then:

$$\frac{\partial f_2}{\partial P_d}\bigg|_{(x_0,u_0)}=\frac{-a^2k_du_{out}^3}{2V_d\sqrt{P_{d_0}-P_{out}}}=a_{22}$$
$$\frac{\partial f_2}{\partial Q_c}\bigg|_{(x_0,u_0)}=\frac{a^2}{V_d}=a_{23} \tag{22}$$

All other partials in the second row of $A$ are zero. If $\dot{Q}_c$ is considered as $f_3$, then:

$$\frac{\partial f_3}{\partial P_s}\bigg|_{(x_0,u_0)}=\frac{A_c}{L_c}\Pi_c(\omega_{c_0},Q_{c_0})=a_{31}$$
$$\frac{\partial f_3}{\partial P_d}\bigg|_{(x_0,u_0)}=-\frac{A_c}{L_c}=a_{32}$$
$$\frac{\partial f_3}{\partial Q_c}\bigg|_{(x_0,u_0)}=P_{s_0}\sum_{i=1}^{3}\sum_{j=0}^{3}i\alpha_{ij}\omega_{c_0}^jQ_{c_0}^{i-1}=a_{33}$$
$$\frac{\partial f_3}{\partial \omega_c}\bigg|_{(x_0,u_0)}=P_{s_0}\sum_{i=0}^{3}\sum_{j=1}^{3}j\alpha_{ij}\omega_{c_0}^{j-1}Q_{c_0}^{i}=a_{34} \tag{23}$$

Finally, if $\dot{\omega}_c$ is considered as $f_4$, then:

$$\frac{\partial f_4}{\partial Q_c}\bigg|_{(x_0,u_0)}=-J^{-1}k_c\omega_{c_0}=a_{43}$$
$$\frac{\partial f_4}{\partial \omega_c}\bigg|_{(x_0,u_0)}=-J^{-1}(k_cQ_{c_0}+2k_l\omega_{c_0})=a_{44} \tag{24}$$

All other partials in the fourth row of $A$ are zero. Eventually, the state matrix $A$ for the EDCS is drawn as follows.

$$A=\begin{bmatrix} a_{11} & 0 & a_{13} & 0 \\ 0 & a_{22} & a_{23} & 0 \\ a_{31} & a_{32} & a_{33} & a_{34} \\ 0 & 0 & a_{43} & a_{44} \end{bmatrix} \tag{25}$$

Moreover, the input derivative is shown in (26). This gives matrix $B$.

$$\frac{\partial f_4}{\partial T_m}\bigg|_{(x_0,u_0)}=J^{-1}=b_{41} \tag{26}$$

All other partials in the remaining rows of matrix $B$ are zero. Hence, the input matrix ($B$) and the out matrix ($C$) of the linearized compressor model are as follows. Matrix $D$ is zero.

$$B=\begin{bmatrix}0 & 0 & 0 & b_{41}\end{bmatrix}^T \tag{27}$$

$$C=\begin{bmatrix}0 & 0 & 1 & 0\end{bmatrix} \tag{28}$$

Finally, the standard controllable state-space realization for the EDCS can be indicated as follows.

$$\begin{bmatrix}\dot{P}_s & \dot{P}_d & \dot{Q}_c & \dot{\omega}_c\end{bmatrix}^T = A\begin{bmatrix}P_s & P_d & Q_c & \omega_c\end{bmatrix}^T + BT_m$$
$$y = C\begin{bmatrix}P_s & P_d & Q_c & \omega_c\end{bmatrix}^T \tag{29}$$

Now, the linearized transfer function of the EDCS can be obtained by forming $C(sI - A)^{-1}B + D$, as shown in (30).

$$G_C(s)=\frac{Q_C(s)}{T_m(s)}=\frac{b_{41}a_{34}(s-a_{11})(s-a_{22})}{s^4+c_3s^3+c_2s^2+c_1s+c_0} \tag{30}$$

where $c_i$ coefficients are calculated based on the $A$ matrix elements, shown in (31).

$$\begin{aligned} c_3 &= -(a_{11}+a_{22}+a_{33}+a_{44}) \\ c_2 &= (a_{11}a_{22}+a_{11}a_{33}+a_{11}a_{44}+a_{22}a_{33}+a_{22}a_{44}+a_{33}a_{44}) \\ &\quad -(a_{13}a_{31}+a_{23}a_{32}+a_{34}a_{43}) \\ c_1 &= (a_{11}a_{23}a_{32}+a_{11}a_{34}a_{43}+a_{22}a_{13}a_{31}+a_{22}a_{34}a_{43}+a_{44}a_{13}a_{31}+ \\ &\quad a_{44}a_{23}a_{32})-(a_{11}a_{22}a_{33}+a_{11}a_{22}a_{44}+a_{11}a_{33}a_{44}+a_{22}a_{33}a_{44}) \\ c_0 &= a_{11}a_{22}a_{33}a_{44}-a_{11}a_{22}a_{34}a_{43}-a_{11}a_{44}a_{23}a_{32}-a_{22}a_{44}a_{13}a_{31} \end{aligned} \tag{31}$$

In the case that the EDCS is controlled, the tracking error is $e(t) = Q_{C\text{-ref}}(t) - Q_C(t)$, as shown in Fig. 3. The closed-loop characteristic equation for the compression system, including the PID controller and the EDCS, is given in (32).

$$\Delta_{cl}(s) = 1 + C_{PID}(s)G_C(s) = 0 \tag{32}$$

Then, the closed-loop characteristic polynomial can be obtained by subsituiting (5) and (30) in (32), as shown in (33). Equation (33) is the key for PID design. In this case, we pick a desired stable fifth-order polynomial, then match coefficients. The desired closed-loop poles are generally chosen such as $s_{1,2,3,4,5} = p_1, p_2, p_3, p_4, p_5$ ($p_i < 0$), as indicated in (34).

$$\begin{aligned}\Delta_{cl}(s) = s^5 &+ (c_3 + K_d b_{41} a_{34}) s^4 \\ &+ (c_2 + b_{41} a_{34}(-K_d(a_{11} + a_{22}) + K_p)) s^3 \\ &+ (c_1 + b_{41} a_{34}(K_d a_{11} a_{22} - K_p(a_{11} + a_{22}) + K_i)) s^2 \\ &+ (c_0 + b_{41} a_{34}(K_p a_{11} a_{22} - K_i(a_{11} + a_{22}))) s \\ &+ (K_i b_{41} a_{34} a_{11} a_{22})\end{aligned} \tag{33}$$

$$\Delta_d(s) = s^5 + d_4 s^4 + d_3 s^3 + d_2 s^2 + d_1 s + d_0 \tag{34}$$

where, coefficients $d_i$ are calculated as follows.

$$\begin{aligned} d_4 &= \sum_{i=1}^{5} p_i \\ d_3 &= \sum_{1 \le i < j \le 5} p_i p_j \\ d_2 &= \sum_{1 \le i < j < k \le 5} p_i p_j p_k \\ d_1 &= \sum_{1 \le i < j < k < l \le 5} p_i p_j p_k p_l \\ d_0 &= \prod_{i=1}^{5} p_i \end{aligned} \tag{35}$$

Then, matching the desired polynomial with the closed-loop characteristic equation obtained in (33) will determine PID gains. This is shown in (36).

$$\begin{aligned} K_i &= \frac{d_4}{b_{41} a_{34} a_{11} a_{22}} \\ K_d &= \frac{d_4 - c_3}{b_{41} a_{34}} \\ K_p &= \frac{d_3 - c_2 - K_d(b_{41} a_{34}(a_{11} + a_{22}))}{b_{41} a_{34}} \end{aligned} \tag{36}$$

It is important to note that, checking degree-of-freedom when matching $\Delta_d(s) = \Delta_{cl}(s)$ is required in this case. We have five equations but only three unknowns ($K_p$, $K_i$, $K_d$). In another words, PID gains can be obtained by using three of the equations, and then the remaining two equations should be checked as consistency constraints. In (36), $K_d$, $K_p$, and $K_i$ are obtained based on $s^4$, $s^3$, and the constant term, respectively. Hence, consistency constraints associated with $s^2$ and $s^1$ terms must be checked, as presented in (37). This ensures the feasibility of the designed PID controller.

$$\begin{aligned} d_2 &= c_1 + b_{41} a_{34}\left(a_{11} a_{22} K_d - (a_{11} + a_{22}) K_p + K_i\right) \\ d_1 &= c_0 + b_{41} a_{34}\left(a_{11} a_{22} K_p - (a_{11} + a_{22}) K_i\right) \end{aligned} \tag{37}$$

Finally, the desired poles are selected based on the closed-loop performance targets and PID gains are calculated accordingly. Fast tracking and conservative responses will be also investigated in this case in Section V.

## IV. Input Data and Simulation Setup

### A. Input Data

This section provides the input data including all parameters and initial conditions considered for the simulation studies. Table I presents the related data to the electrolyzer plant and EDCS. The nominal active power of the electrolyzer plant and the EDCS are 5.2 GW and 97.4 MW, respectively. At these power levels, the electrolyzer produces a constant hydrogen flow of 32.5 kg/s, while the EDCS raises the outlet pressure to the desired setpoint. The EDCS driver power consumption (kW) is computed using (38), as described in [27].

$$P_{\mathrm{D}} = 4.276 Z_{\mathrm{ave}} \frac{Q_{\mathrm{c}} T_{\mathrm{s}}}{E_{\mathrm{c}} \eta_{\mathrm{c}} \eta_{\mathrm{m}}} \frac{\kappa}{\kappa - 1} \left[ \left(\frac{P_{\mathrm{d}}}{P_{\mathrm{s}}}\right)^{\frac{\kappa-1}{\kappa}} - 1 \right] \tag{38}$$

where, $Z_{ave}$ is the average compressibility factor of $H_2$. $E_c$, $\eta_c$, and $\eta_m$ are parasitic efficiency, compressor efficiency, and mechanical efficiency of the compressor driver, respectively; EDCS's flow rate, inlet temperature, inlet pressure, and outlet pressure are shown by $Q_c$, $T_s$, $P_s$, and $P_d$, respectively; and κ is the ratio of specific heats of $H_2$.

TABLE I
Numerical Parameters and Initial Conditions Used for the Investigated EDCS.

| | | Parameters | Values [unit] |
|---|---|---|---|
| EDCS | Compressor Specifications | $V_s$ | 0.2 m$^3$ |
| | | $V_d$ | 0.1 m$^3$ |
| | | a | 340 m/s |
| | | $A_c$ / $L_c$ | 1×10$^{-4}$ m |
| | | J | 0.1 kgm$^2$ |
| | | $k_s$ | 0.3 |
| | | $k_d$ | 0.5 |
| | | $k_c$ | 0.002 |
| | | $k_l$ | 0.002 |
| | | $Z_{ave}$ | 1.06 |
| | | $E_c$ | 0.99 |
| | | $\eta_c$ | 0.87 |
| | | $\eta_m$ | 0.95 |
| | | $T_s$ | 288.14 K |
| | | κ | 1.41 |
| | Boundary Conditions | $P_{in}$ | 30.5 bar-g |
| | | $P_{out}$ | 115 bar-g |
| | Valve Positions | $u_{in}$ | 60 % |
| | | $u_{out}$ | 40 % |
| | Initial Values | $P_{s0}$ | 28 bar-g |
| | | $P_{d0}$ | 125.4 bar-g |
| | | $Q_{c0}$ | 32.65 kg/s |
| | | $\omega_{c0}$ | 4984 rad/min |
| Electrolyzer | Electrolyzer Specifications | $V_{ES}$ | 40 V |
| | | α | 123.8 |
| | | β | 61.1 |
| | | $\eta_f$ | 1 |
| | | $N_s$ | 24 |
| | | $N_p$ | 1.3×10$^6$ |

Additionally, the compressor map is approximated by a third-order polynomial in $\omega_c$ and $Q_c$, see (17). In this regard, Table II lists the corresponding $\alpha_{ij}$ coefficients to this map.

TABLE II
Coefficients of the EDCS Map.

| | $\omega_c^0$ | $\omega_c^1$ | $\omega_c^2$ | $\omega_c^3$ |
|---|---|---|---|---|
| $Q_c^0$ | 3.9093 | 1.088×10$^{-4}$ | 7.8014×10$^{-9}$ | 1.6728×10$^{-13}$ |
| $Q_c^1$ | 1.4021×10$^{-2}$ | 2.7584×10$^{-6}$ | -1.3803×10$^{-10}$ | 1.2765×10$^{-14}$ |
| $Q_c^2$ | -8.3373×10$^{-4}$ | -1.106×10$^{-7}$ | 1.7853×10$^{-11}$ | -1.093×10$^{-15}$ |
| $Q_c^3$ | 1.1547×10$^{-6}$ | 1.8473×10$^{-9}$ | -4.1134×10$^{-13}$ | 2.4067×10$^{-17}$ |

### B. Simulation Setup

The integrated electrolyzer-EDCS system is implemented in MATLAB. Four case studies are considered to analyze the dynamic behavior of the system considered. Cases 1 and 2 represent scenarios where the EDCS and the electrolyzer face power grid disturbances, respectively, without any control link between them. Cases 3 and 4 correspond to the implementation of PID controllers for Cases 1 and 2, respectively.

*Case 1*: System response to power grid disturbances on the EDCS without control.

In Case 1, a power disturbance is applied to the EDCS driver. The resulting impact on the system dynamics is then evaluated to illustrate the system response in the absence of a feedback controller for mitigating driver power deviations. The imposed driver power deviation is depicted in Fig. 4.

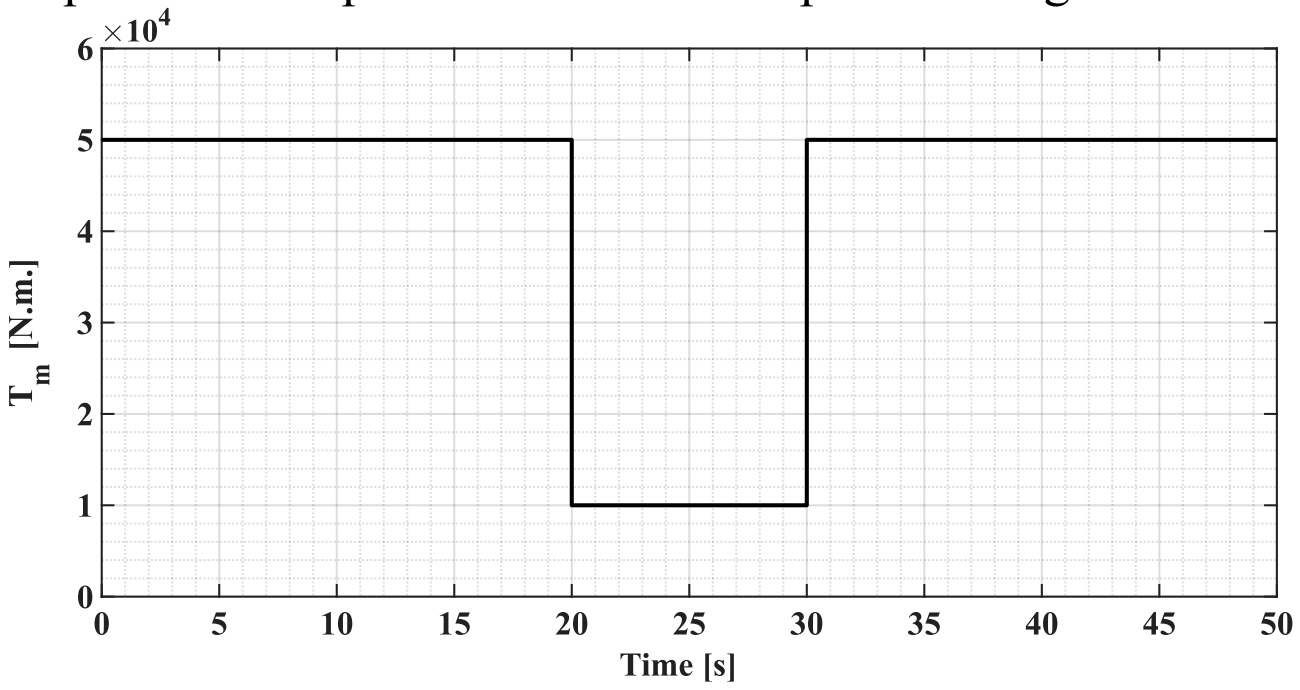

**Fig. 4.** Power grid disturbance on the EDCS driver – Case 1.

*Case 2*: System response to power grid disturbances on the electrolyzer without control.

In Case 2, the electrolyzer faces power disturbance. This disturbance applied to the input voltage of the electrolyzer stack, resulting in deviations in the $H_2$ production. Figure 5 demonstrates the resulting variations in the electrolyzer outlet $H_2$ flow rate. Then, the impact on the system dynamics is analyzed to observe the system response in the absence of a proper control mechanism.

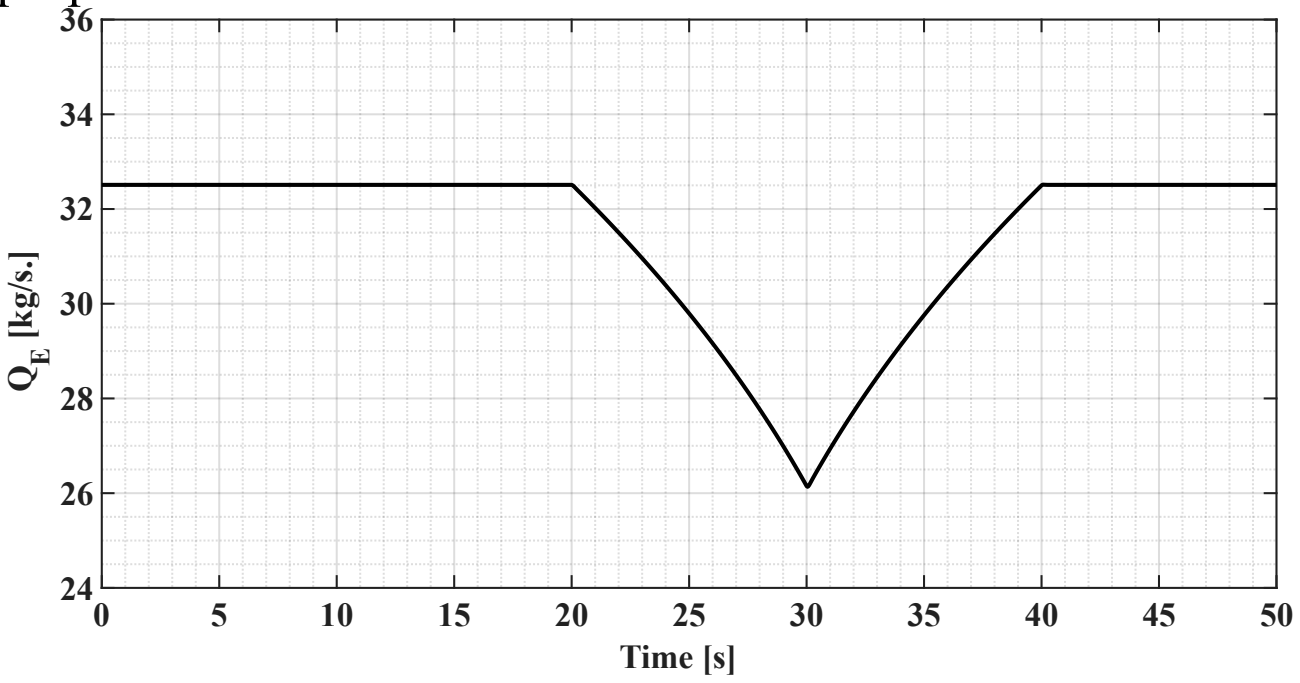

**Fig. 5.** Electrolyzer flow deviations – Case 2.

*Case 3*: Controller implementation for the electrolyzer when the EDCS is imposed by power grid disturbance.

In Case 3, a power disturbance similar to Case 1 is applied on the compressor station. But in this case, the designed PID controller is installed to adjust the electrolyzer operating point based on the EDCS driver deviations. This control mechanism would guarantee the safe operation of the system and reduce the risk of equipment damage. It is worth mentioning that inadequate coordination between the electrolyzer and the compressor under these conditions could create hazardous fluctuations in compressor operation, which may compromise reliability [27].

*Case 4*: Controller implementation for the EDCS when the electrolyzer is imposed by power grid disturbance.

In Case 4, a power disturbance similar to Case 2 is applied on the electrolyzer. However, a PID controller is installed to adjust the dynamic operation of the EDCS in response to these $H_2$ production deviations. This control mechanism would also be intended to coordinate the operation of the electrolyzer and the compressor and to maintain the safe operation of the system. Notably, ramp-shaped disturbance is considered at the electrolyzer outlet, as it more closely reflects realistic gas network dynamics [4], [31]. The proposed control scheme, however, is not limited to this disturbance type and can be extended to other disturbance profiles.

## V. Simulation Results

This section presents and analyzes the simulation outcomes obtained in this study. First, the system responses in Cases 1 and 2 are examined to highlight the need for appropriate control links between the electrolyzer and the EDCS. Then, the results in Cases 3 and 4 are discussed to demonstrate coordinated system operation achieved with the design of PID controllers.

Figure 6 demonstrates the driver power of the EDCS during power grid disturbance. Prior to the driver contingency, the EDCS power consumption ($P_D$) is 97.87 MW, which decreases to 86.7 MW during the torque loss. This consumption drop is observed with a significant fluctuation at the beginning and the end of the disturbance. This affects all system variables.

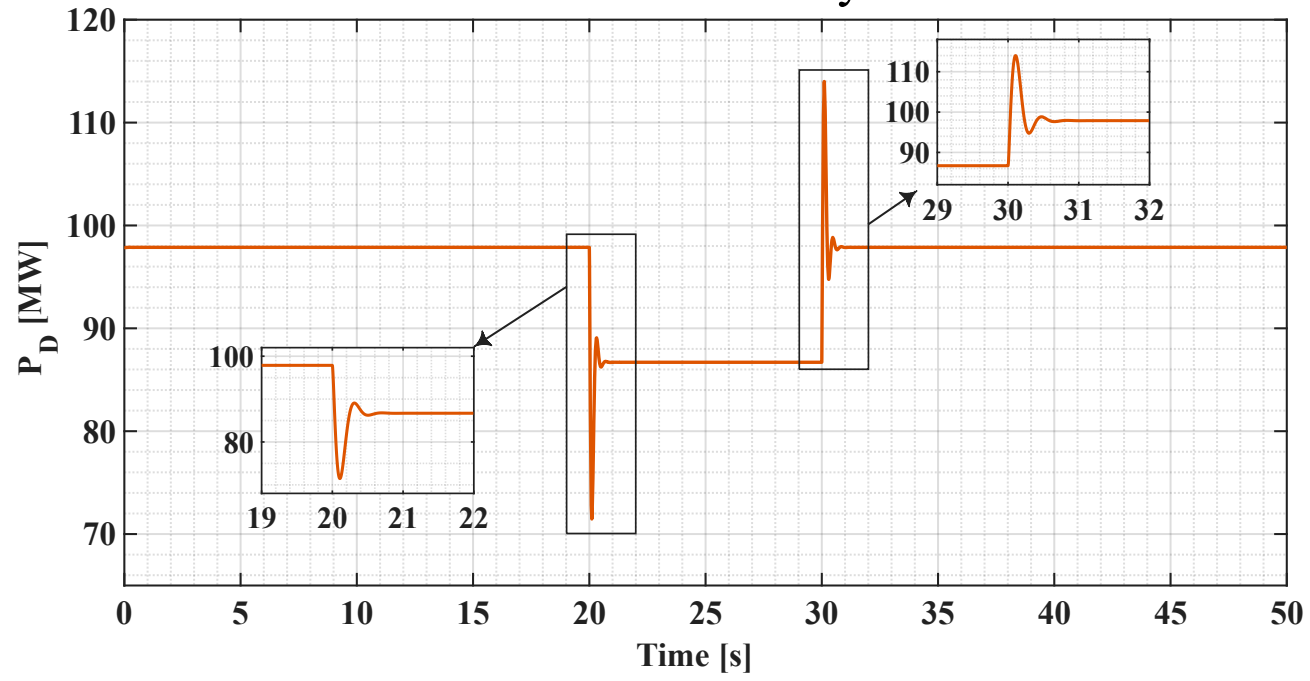

**Fig. 6.** Driver power of the EDCS during motor torque disturbance – Case 1.

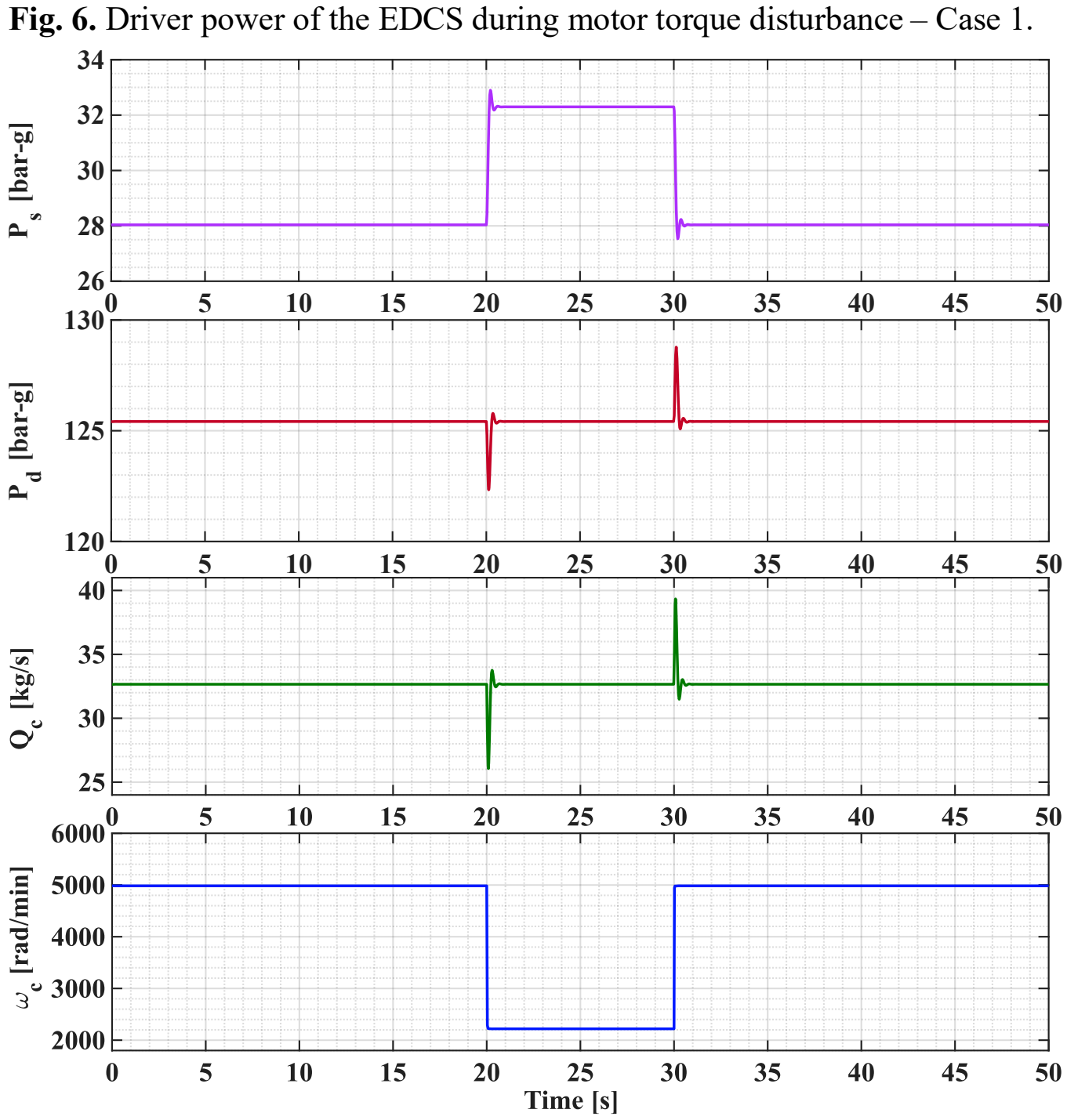

**Fig. 7.** State variables of the EDCS during motor torque disturbance – Case 1.

Dynamics of the EDCS's suction pressure ($P_s$), discharge pressure ($P_d$), flow rate ($Q_c$), and rotational speed ($\omega_c$) during the disturbance are depicted in Fig. 7. $\omega_c$ drastically drops, then it recovers to its steady-state value after the contingency. In contrast, the gas variables behave differently. $Q_c$ and $P_d$ initially decrease during the disturbance but quickly recover due to the constant input flow. As the driver torque recovers, a peak in $Q_c$ and $P_d$ is observed (see 30-31 *sec* in Fig. 7). However, $P_s$ behaves differently, and rises to a new level during the disturbance, as explained in [27].

Figure 8 stands for the dynamics of the EDCS state variables under electrolyzer contingency. When $Q_E$, which is also the suction flow ($Q_s$) of the EDCS, is dropped, $Q_c$, $P_s$, and $P_d$ show similar behaviors and decrease accordingly. On the other hand, $\omega_c$ just increases slightly during the contingency as the motor torque ($T_m$) has remained constant at its nominal value ($5\times10^4$ N.m.). Also, under this condition, $P_D$ decreases from 97.87 MW to 81.5 MW, which is proportional to the reductions of $Q_c$, $P_s$, and $P_d$ based on (37). $P_D$ of the EDCS is demonstrated in Fig. 9.

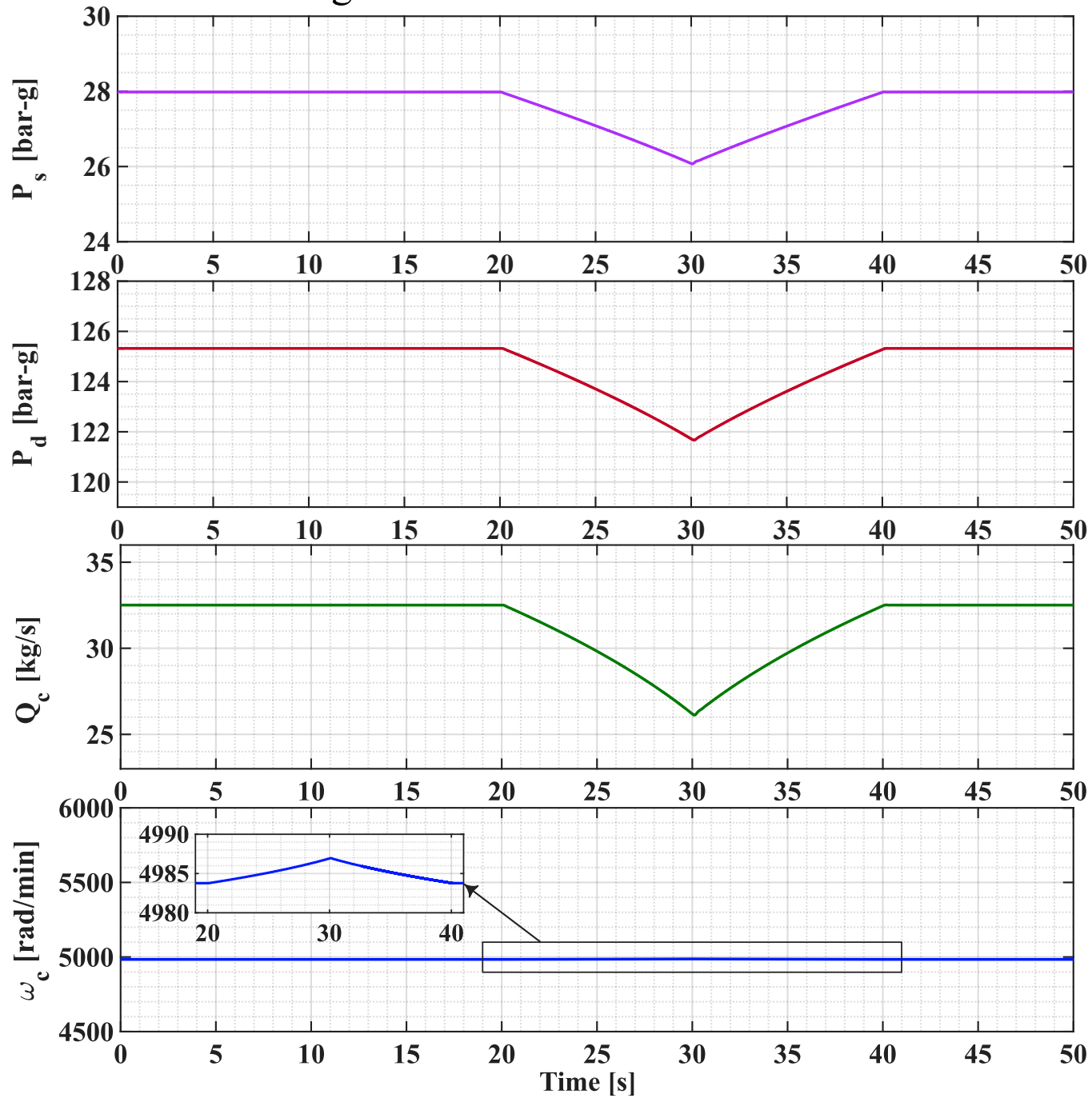

**Fig. 8.** State variables of the EDCS during electrolyzer deviations – Case 2.

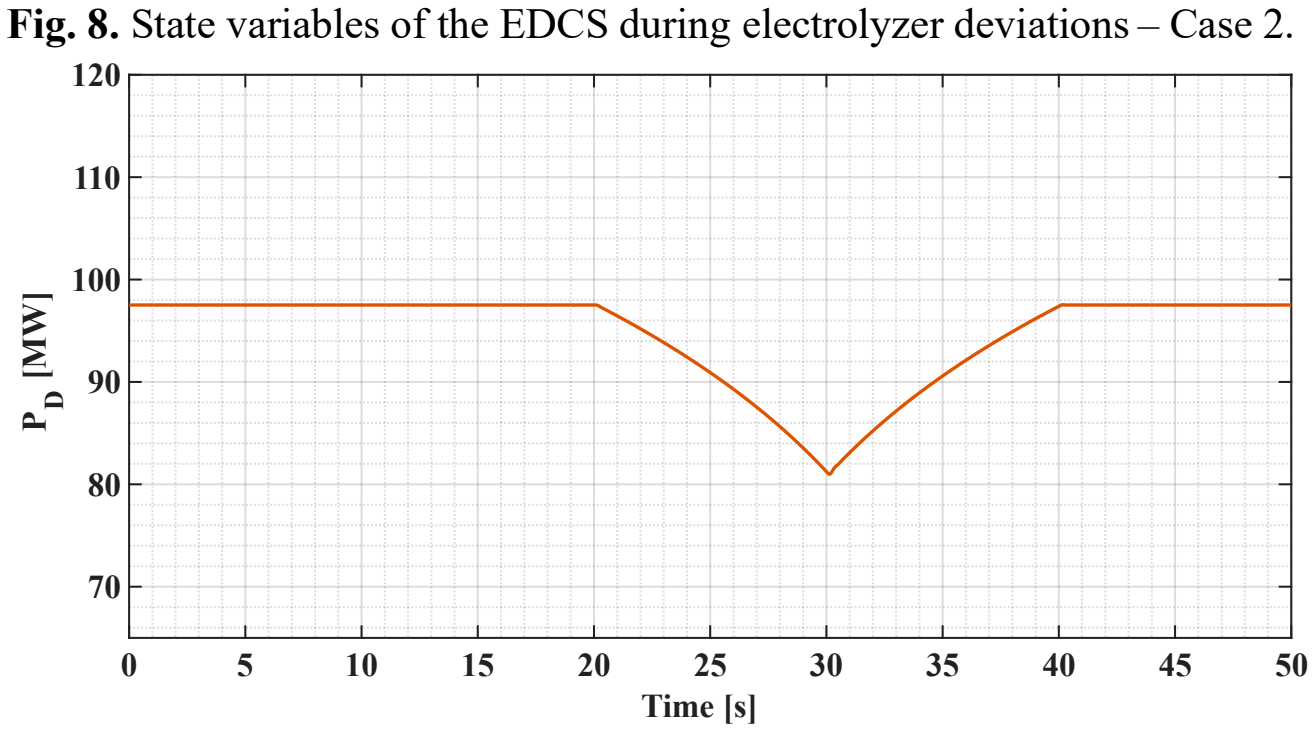

**Fig. 9.** Driver power of the EDCS during electrolyzer deviations – Case 2.

To facilitate a better understanding of the EDCS dynamics in Cases 1 and 2, the behaviors of the state variables in these cases are summarized as follows. The normal expected dynamic behavior of the compressor variables and $Q_E$ during compressor torque drop is presented in (39) [4].

$$T_m \downarrow \;\Rightarrow\; \omega_c,\; Q_E \downarrow \;\Rightarrow\; Q_c \downarrow \;\Rightarrow\; P_s \uparrow,\; P_d \downarrow \tag{39}$$

However, observed dynamics in Case 1 are presented in (40).

$$T_m \downarrow \;\Rightarrow\; \omega_c \downarrow,\;\; Q_c, Q_E = \text{const.},\;\; P_d = \text{const.},\;\; P_s \uparrow \tag{40}$$

Also, the normal expected dynamics of the EDCS variables and $T_m$ during electrolyzer flow deviations are presented in (41).

$$Q_E \downarrow \;\Rightarrow \begin{cases} Q_c \downarrow \;\Rightarrow\; \omega_c \downarrow \;\Rightarrow\; P_s \uparrow,\; P_d \downarrow \\ T_m \downarrow \end{cases} \tag{41}$$

However, observed dynamics in Case 2 are presented in (42).

$$Q_E \downarrow \;\Rightarrow\; Q_c \downarrow,\; P_s \downarrow,\; P_d \downarrow,\; \omega_c, T_m \approx \text{const.} \tag{42}$$

Based on the conceptual compressor map, shown in Fig. 10, both dynamic trajectories in Cases 1 and 2 can be justified theoretically by using (10)-(13). In Fig. 10, the normal path presents dynamics seen in (39) and (41), while paths marked for Cases 1 and 2 shows dynamics in (40) and (42), respectively. Also, the operation trajectory of the EDCS in Case 1 (see Fig. 6) and Case 2 (see Fig. 8) is depicted in Fig. 11. Besides abnormal operation in both cases, the EDCS approaches the choke line during the disturbance that happened in Case 1. This exactly corresponds to the moments that the compressor variables experience fluctuations with a large magnitude at the beginning and end of the event. This behavior is undesirable from the safety perspective [32], [33].

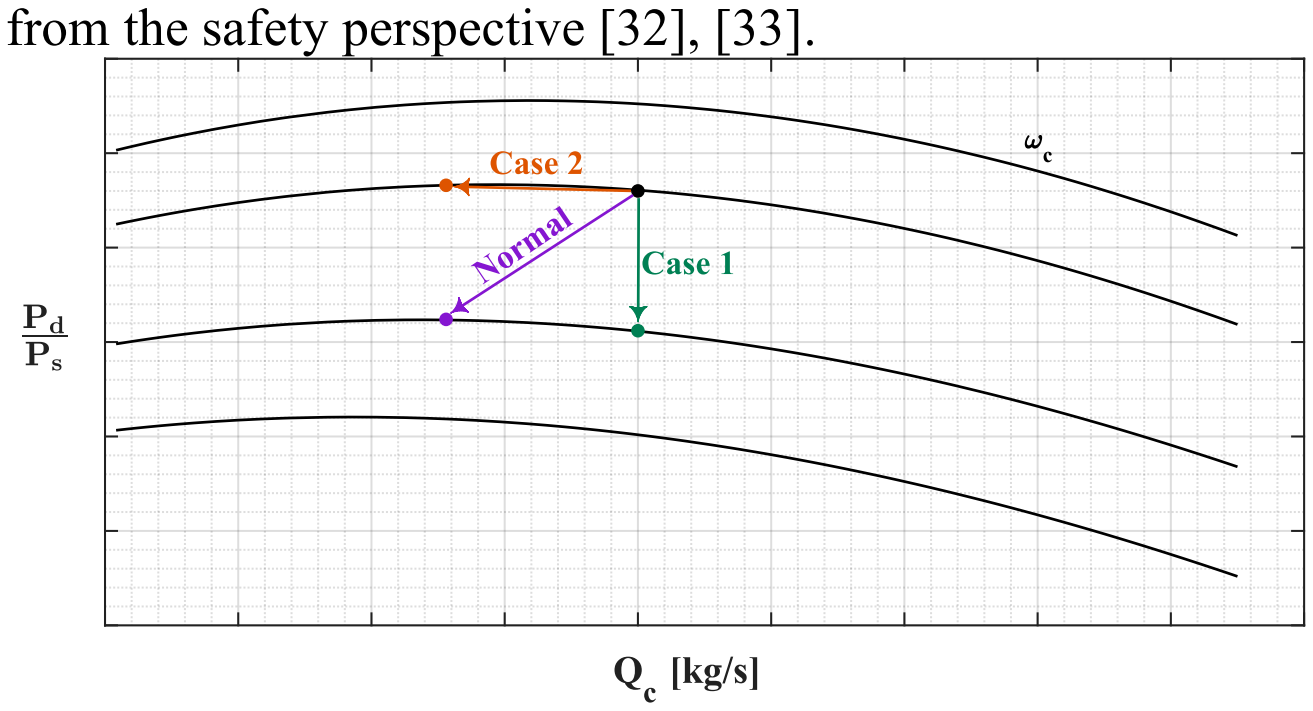

**Fig. 10.** Normal and abnormal dynamic operation of the EDCS under motor torque and electrolyzer contingencies.

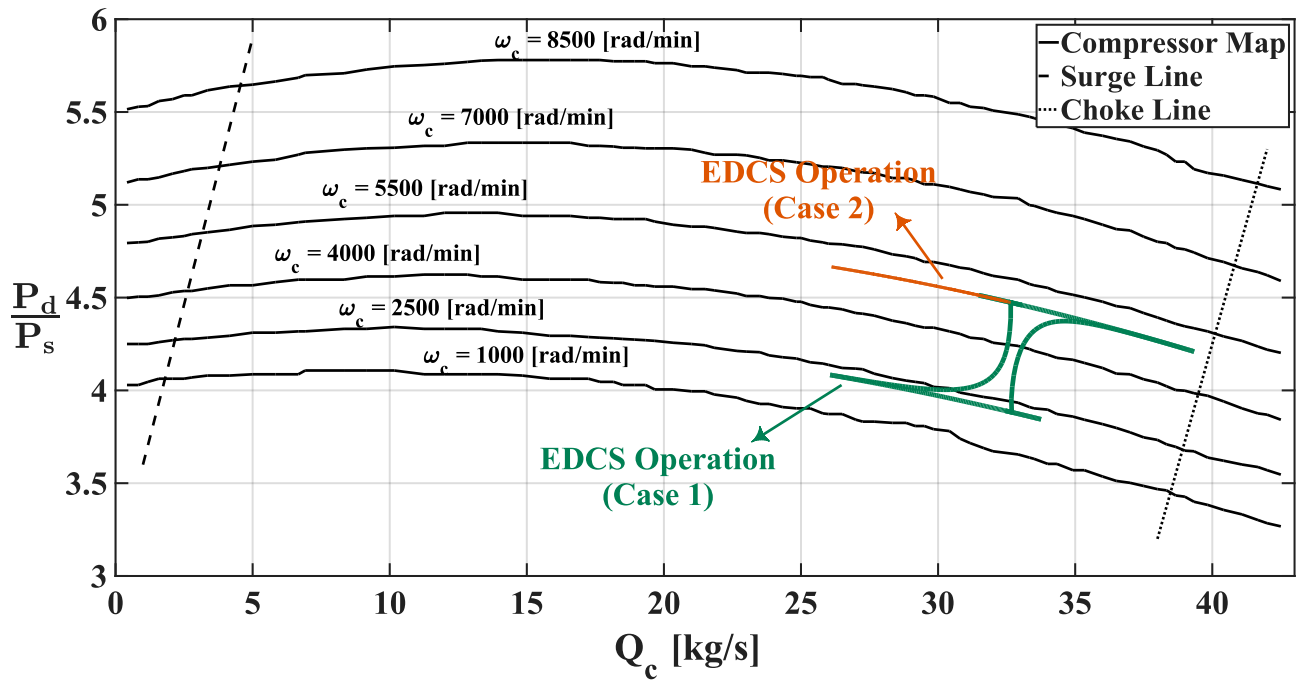

**Fig. 11.** Operation trajectory of the EDCS in Cases 1 and 2.

Overall, dynamic behaviors observed in Cases 1 and 2 indicate inconsistencies in system response. In real-world operations in Case 1, all state variables of the compressor station should change accordingly when a driver power loss occurs. For instance, the compressor flow should decrease, and this reduction implies that the electrolyzer should also decrease its $H_2$ production. Also, when the electrolyzer flow ($=Q_s=Q_c$) is disturbed in Case 2, the compressor driver should control its torque accordingly. This prevents any damage to the compressor's internal equipment. For instance, when $Q_c$ decreases, the compressor should decrease its torque, resulting in reducing $\omega_c$ with a proper control action. However, these actions do not occur in Cases 1 and 2 due to the absence of a proper control link between the electrolyzer and the EDCS.

This highlights the need for addressing the existing gap in the literature and developing an integrated control strategy for the coupled electrolyzer–compressor system.

In Case 3, we investigate the necessity of the electrolyzer to regulate its output flow in response to the disturbance applied to the EDCS. To this end, two PID controllers are designed for $Q_E$: one to provide a conservative response and the other to provide a fast-tracking response. Table III lists the PID coefficients used in Case 3. For the conservative design, the desired closed-looped poles are chosen as $s_1 = -1.5$, $s_2 = -60$ and $s_3 = -125$, with the corresponding PID gains obtained from (8) and (9). Here, $s_1$ is the dominant pole and behaves like a slow integrator, reducing bandwidth and improving robustness to modeling uncertainty and noise. The remaining poles, $s_2$ and $s_3$, are placed close to the open-loop poles α and β, meaning that the PID controller does not significantly alter their dynamics. Overall, the dominant pole is selected to satisfy the desired settling time, estimated as $t_s = 4/|s_1|$, while the other poles are placed sufficiently far from it to remain non-dominant and to avoid affecting the primary closed-loop response. For the fast design, the desired closed-looped poles are chosen as $s_1 = -50$, $s_2 = -100$ and $s_3 = -200$, with the PID gains obtained from (8) and (9). All three poles are selected well beyond the imaginary axis with a dominant pole at s = −50 and a much wider closed-loop bandwidth.

TABLE III
PID COEFFICIENTS DESIGNED FOR THE ELECTROLYZER.

| Conservative Response | Value | Fast-tracking Response | Value |
|---|---|---|---|
| $\mathbf{K_p}$ | 0.086 | $\mathbf{K_p}$ | 11.13 |
| $\mathbf{K_i}$ | 4.56 | $\mathbf{K_i}$ | 405.58 |
| $\mathbf{K_d}$ | $6 \times 10^{-4}$ | $\mathbf{K_d}$ | 0.067 |

After applying the designed controller to the system, the controlled flow rate of the electrolyzer, $Q_E$, can be traced in Fig. 12. The figure compares $Q_E$ for the case without a control link between the electrolyzer and the EDCS (Case 1) and for the cases in which PID controllers are designed to provide conservative and fast responses. As shown, the electrolyzer effectively reduces its output flow in response to the disturbance affecting the EDCS driver. In addition, the hazardous undershoot and overshoot transients observed in Case 1 are eliminated by the PID implementation. It is worth mentioning that both controllers achieve satisfactory reference tracking under nominal conditions. The conservative design uses dominant poles closer to the imaginary axis, resulting in a lower closed-loop bandwidth and a slower response. In contrast, the fast-tracking design employs poles farther into the left-half plane, leading to a higher closed-loop bandwidth and faster nominal tracking.

Accordingly, the inconsistencies in the EDCS dynamics are resolved when $Q_E$ is properly controlled, as shown in Figs. 13 and 14. All EDCS state variables exhibit the expected behavior under motor disturbance, consistent with (39). In addition, the hazardous fluctuations in the compressor driver power are eliminated, and the power consumption during the disturbance is correctly reduced from 97.87 MW to 67 MW (compare Figs. 7 and 14). This consumption level is the actual expected driver power of the EDCS during the disturbance. Furthermore, the operation trajectory of the EDCS after controlling $Q_E$ is demonstrated in Fig. 15. In this case, the compressor operates normally and does not approach the choke line.

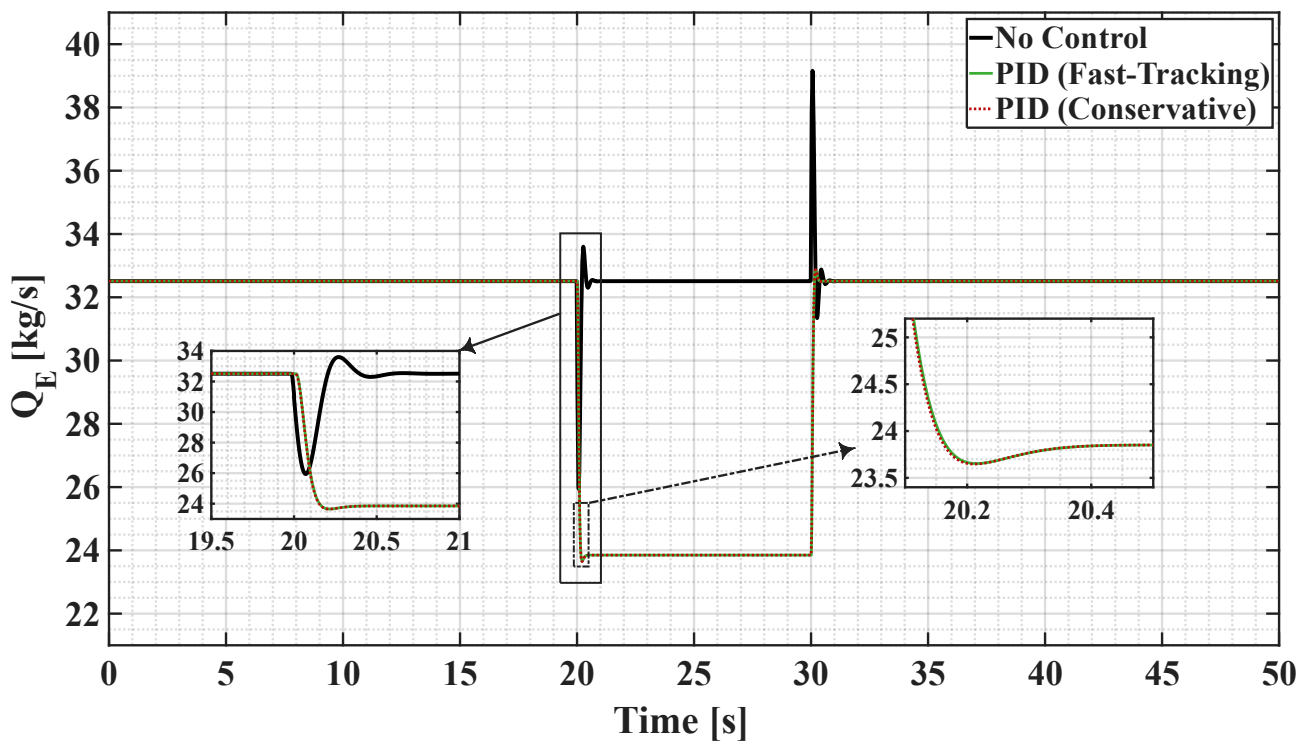

**Fig. 12.** Electrolyzer flow rate with and without the PID – Case 3.

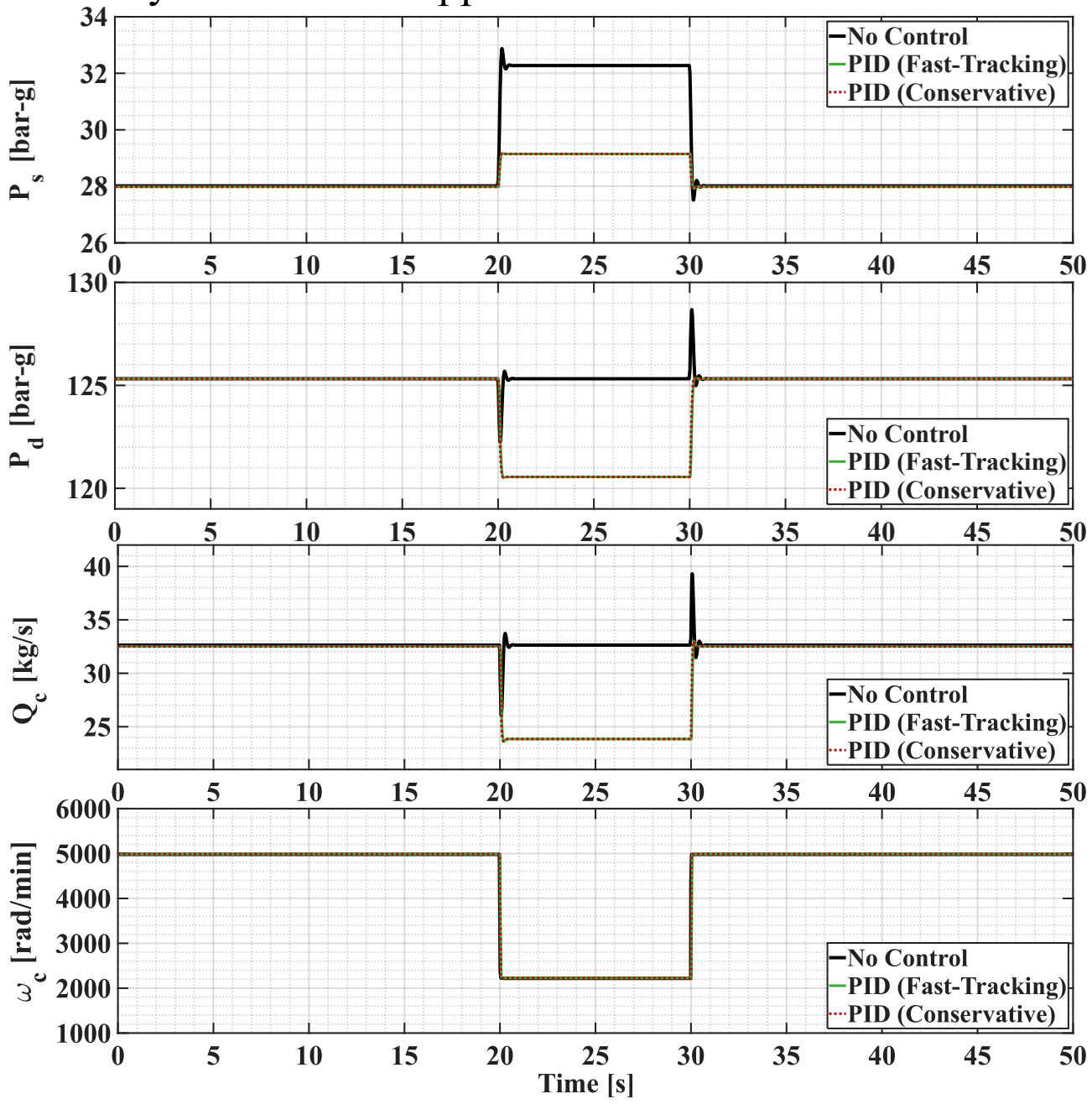

**Fig. 13.** State variables of the EDCS after applying the PID – Case 3.

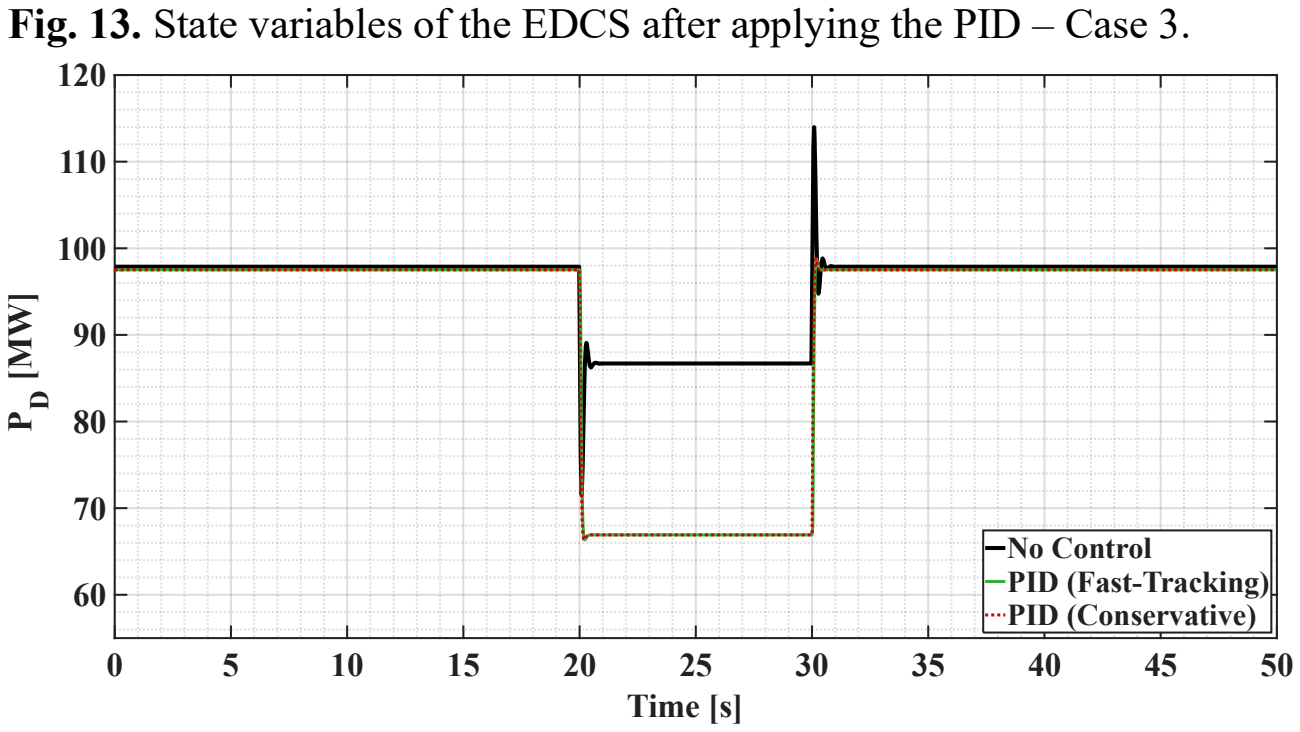

**Fig. 14.** Driver power of the EDCS after applying the PID – Case 3.

In Case 4, the EDCS is required to control its torque in response to the disturbance applied to the electrolyzer. Similar to Case 3, two PID controllers with conservative and fast-tracking responses are designed for controlling $T_m$. The obtained PID gains for this case are listed in Table IV. The desired closed-loop poles were selected as $s = [-1, -20, -50, -100, -200]$ for the conservative design. Using the pole placement in (36), the PID gains are computed as given in Table IV. However, enforcing the consistency constraints in (37) with the computed PID coefficients, the resulting actual closed-loop poles are located at $s$ = [−0.77, −15.03, −64.38 ± 58.69j, −226.44], which still indicates a stable closed-loop response. For the fast-tracking design, the desired closed-looped poles were chosen as $s$ = [−8, −40, −100, −200, −380]. After computing the PID gains and applying the consistency constraints, the actual closed-loop poles are obtained as $s$ = [−12.61, −27.84, −101.92 ± 76.11j, −418.73], again confirming stable closed-loop operation. Unlike small laboratory-scale compressor systems, the present study considers a large industrial compressor station in the 100 MW range. Therefore, the large PID gains obtained for the EDCS are consistent with the scale of the system, where the driver torque input operates in the range of $10^4$–$10^5$ N.m., and the compressor flow output is 32 kg/s. Accordingly, the controller must map a relatively small flow-rate error into a large torque correction. This scaling is also consistent with prior studies, where gains on the order of $10^4$ have been reported even for significantly smaller compressor systems [26].

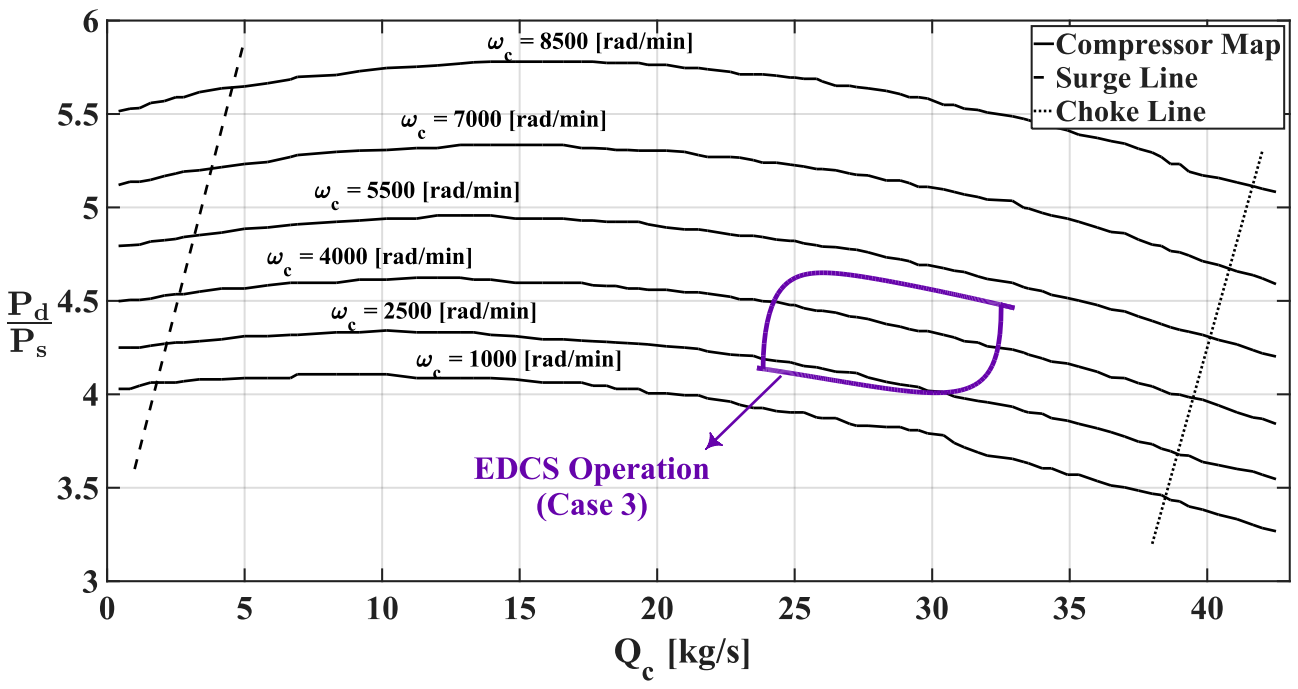

**Fig. 15.** Operation trajectory of the EDCS after applying the PID – Case 3.

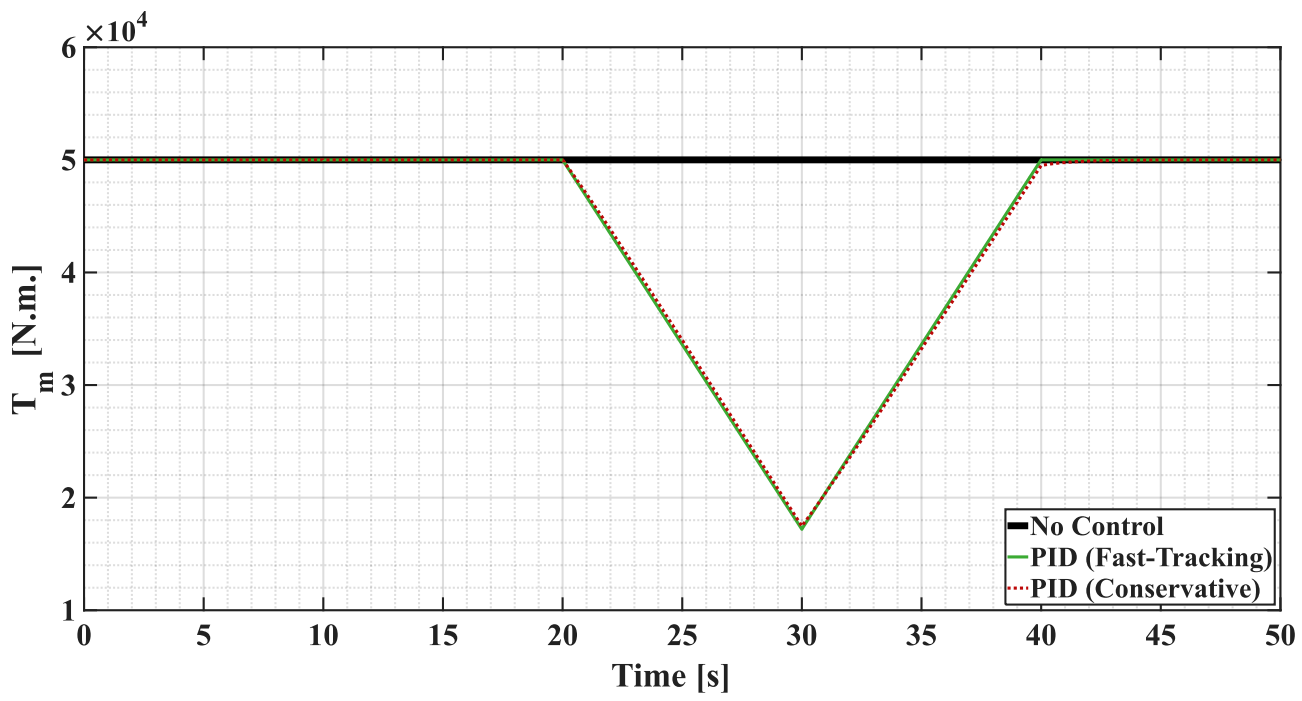

**Fig. 16.** Motor torque of the EDCS with and without the PID – Case 4.

TABLE IV
PID COEFFICIENTS DESIGNED FOR THE EDCS.

| Conservative Response | Value | Fast-tracking Response | Value |
|---|---|---|---|
| $K_p$ | $4.9 \times 10^4$ | $K_p$ | $2.6 \times 10^5$ |
| $K_i$ | $4.3 \times 10^4$ | $K_i$ | $5.2 \times 10^6$ |
| $K_d$ | 155 | $K_d$ | 677 |

Figure 16 illustrates the driver torque of the EDCS. It compares $T_m$ in the uncontrolled case (Case 2), in which no control link exists between the electrolyzer and the EDCS, with the case employing the conservative and fast-tracking PID controllers (Case 4). In the absence of this control link, $T_m$ remains unchanged despite the disturbance applied to $Q_E$. By contrast, once the PID controller is implemented, the EDCS adjusts its driver torque appropriately in response to the electrolyzer flow disturbance. This response confirms the effectiveness of the proposed control scheme in establishing the required dynamic interaction between the electrolyzer and the EDCS. Moreover, both the conservative and fast-tracking designs provide satisfactory reference-tracking performance.

By properly controlling $T_m$, the undesirable EDCS behavior identified in Case 2 is resolved, as indicated in Figs. 17 and 18. The state variables exhibit the expected dynamics under electrolyzer flow disturbance, following (41). The compressor driver power during the disturbance decreases from 97.87 MW to 74.5 MW (compare Figs. 9 and 18). This reduced level corresponds to the expected driver power of the EDCS under this operating condition. The operating trajectory of the EDCS is shown in Fig. 19, confirming acceptable performance for the compressor. The operating point stays mainly within the zoomed area, undergoes a temporary deviation from the base operating zone, and then returns, all without approaching the choke line.

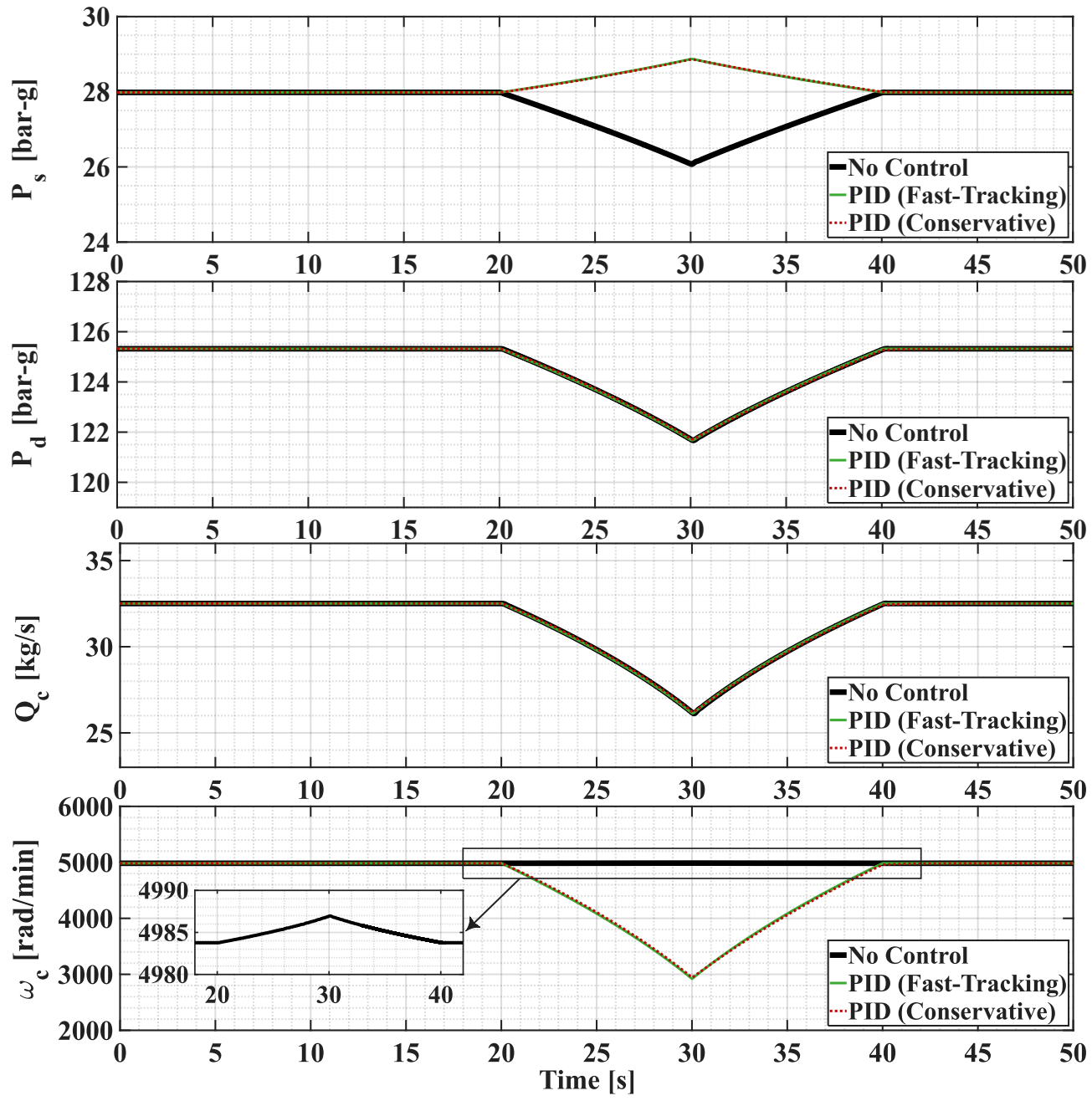

**Fig. 17.** State variables of the EDCS after applying the PID – Case 4.

**Fig. 18.** Driver power of the EDCS after applying the PID – Case 4.

## VI. Conclusion

This work investigates coordinated control of an integrated electrolyzer and electric-driven compressor station (EDCS) system under disturbances occurring to either component. In this regard, an integrated dynamic model has been developed to replicate the dynamic interactions between the electrolyzer and EDCS. Subsequently, it utilizes an existing linear electrolyzer model and develops a linearized model for the EDCS to design conservative and fast-tracking PID controllers.

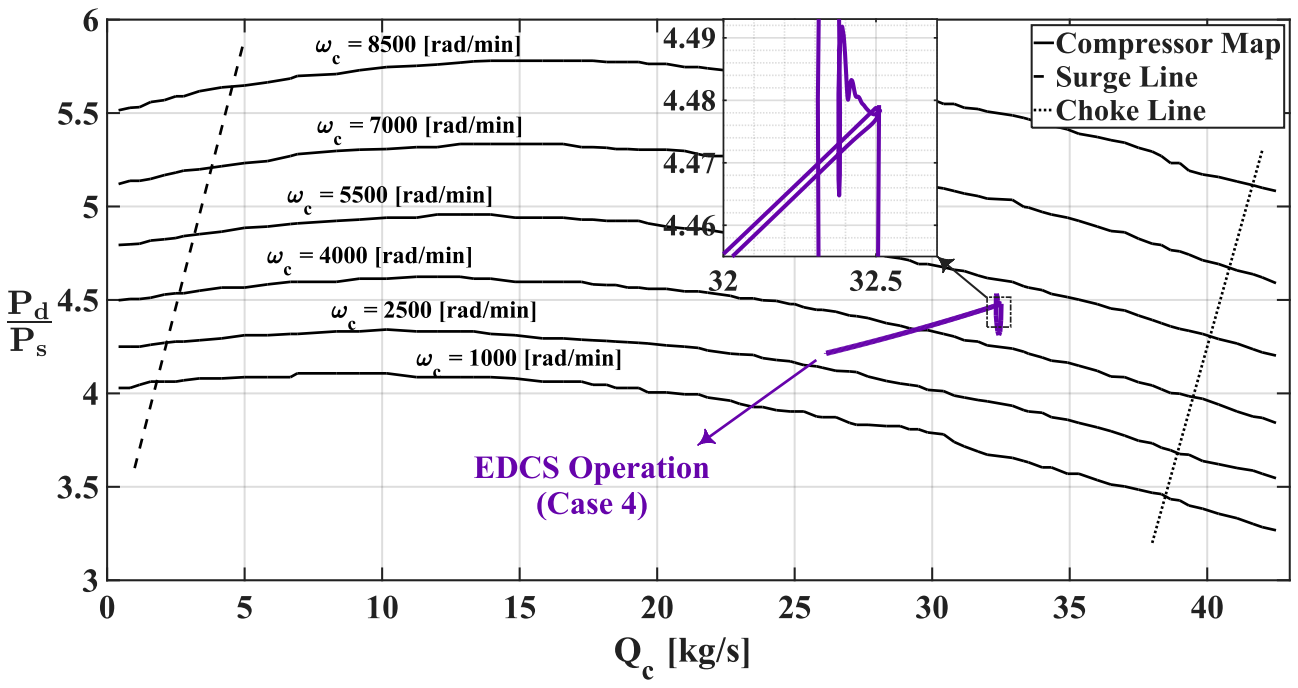

**Fig. 19.** Operation trajectory of the EDCS after applying the PID – Case 4.

Simulation results showed that a mismatch dynamic response will occur under the absence of a coordination link between the electrolyzer and the EDCS. In particular, when a disturbance happens to the EDCS driver while the electrolyzer flow remains unchanged, undesirable pressure and flow transients were observed in the system, and the EDCS approached the choke region. Conversely, an electrolyzer flow disturbance with an unregulated EDCS driver torque results in inconsistent EDCS dynamic responses.

The proposed coordination strategy addressed these negative impacts in both disturbance scenarios. Regulating the electrolyzer hydrogen flow in response to the EDCS driver disturbance, eliminated the observed transient fluctuations, and maintained the compressor away from the choke line. Similarly, controlling the EDCS driver torque in response to the flow disturbances of the electrolyzer, restored the expected dynamic responses in the pressure, flow, and mechanical speed of the EDCS. Also, both conservative and fast-tracking controllers revealed satisfactory tracking performance suitable for practical purposes.